# Multi-Unit Auctions: Beyond Roberts

Shahar Dobzinski
Department of Computer Science
Cornell Unversity
shahar@cs.cornell.edu

Noam Nisan *
School of Computer Science and Engineering
Hebrew University
noam@cs.huji.ac.il

November 2, 2018

**Abstract**

We exhibit incentive compatible multi-unit auctions that are not affine maximizers (i.e., are not of the VCG family) and yet approximate the social welfare to within a factor of $1 + \epsilon$. For the case of two-item two-bidder auctions we show that these auctions, termed Triage auctions, are the only scalable ones that give an approximation factor better than 2. "Scalable" means that the allocation does not depend on the units in which the valuations are measured. We deduce from this that any scalable computationally-efficient incentive-compatible auction for $m$ items and $n \geq 2$ bidders cannot approximate the social welfare to within a factor better than 2. This is in contrast to arbitrarily good approximations that can be reached under computational constraints alone, and in contrast to the fact that the optimal social welfare can be obtained under incentive constraints alone.

*Supported by a grant from the Israeli Academy of Sciences.

# 1 Introduction

**Background**

The field of *Algorithmic Mechanism Design* [27] designs mechanisms for achieving various computational goals, under the assumption of rational selfishness of the involved parties. The notions used are taken from the economic field of Mechanism Design, and a basic notion is that of *incentive-compatibility*– where rational players are motivated to act *truthfully*. For background and survey see part II of [28]. This paper will consider only the simplest and most robust notion of incentive compatibility, that of dominant strategies in quasi-linear settings with independent private values. The typical question in the field asks for a computationally-efficient incentive compatible mechanism that implements a certain type of outcome, usually the approximate optimization of some target "social" goal. There are two variants of this challenge, the first considers situations where incentive compatibility itself is hard to achieve and the computational efficiency is just an additional burden, with the prime example being approximate minimization of the makespan in scheduling problems [27]. The second variant focuses on cases where each of the two constraints of incentive compatibility and computational efficiency can be achieved separately, and the challenge is to get them simultaneously, with the prime example being approximate welfare maximization in various types of combinatorial auctions [24].

While there has been much work and some progress on these types of challenges, with particular emphasis on the problems mentioned above of combinatorial auctions (e.g., [22, 20, 3, 15, 23, 16, 13, 6]) and scheduling (e.g., [7, 21, 2]), the basic challenge remains mostly unanswered. As noted in [22], the main issue turns out to be the richness of the domain of player's valuations, i.e., of their private information. On one extreme are single-dimensional domains where the private information of each participant is captured by a scalar (or domains very close to it, e.g., [24]). For these types of problems, incentive-compatible mechanisms are well characterized by a certain monotonicity condition and, in most cases, the challenge of reconciling incentives with computational efficiency has been met [24, 1, 5, 9, 8]. On the other extreme are problems which are "fully dimensional" (or close to fully dimensional, e.g., [29, 17]) where there is no structure on valuations, in which case a key theorem of Roberts [30] characterizes incentive compatible mechanisms as "affine maximizers" "on a sub-range" – simple generalizations of the VCG mechanism. While such affine maximizers on a sub-range are not completely powerless in polynomial time, in most cases this characterization implies impossibility of good computationally efficient truthful mechanisms. Most interesting problems, including those mentioned above, lie in an intermediate range where the valuation spaces are neither single dimensional nor fully dimensional, a range for which very little is known. The main problem seems to be the lack of a good characterization of incentive compatibility in these intermediate ranges[1]. In particular, the key unknown is whether any useful truthful non-VCG mechanisms exist in the intermediate range[2].

**Multi-unit Auctions**

As mentioned, the paradigmatic problems for the reconciliation of computational constraints with incentive constraints are the various subclasses of combinatorial auctions. In this paper we consider the simplest variant which exhibits this tension: multi-unit auctions. In this problem there are $m$ identical items for sale among $n$ bidders, where each bidder $i$ has a valuation function $v_i : \{0...m\} \to \Re$, where $v_i(k)$ denotes player $i$'s value for receiving $k$ items. The valuations $v_i$ are assumed to be monotone non-decreasing (free disposal) with $v_i(0) = 0$ (normalization). Key and implicit here is that there are no externalities: the value of bidder $i$ depends only on what he gets rather than also on the allocation to the others. The optimization goal is to find an allocation of items to the bidders, where bidder $i$

---
[1] The "weak monotonicity" [4, 31] characterization is from the point of view of a single player and is not specific enough to be useful in this regard.

[2] With a single positive exception for certain multi-unit combinatorial auctions [3].

gets $s_i$ items, with $\sum_i s_i \leq m$, that maximizes social welfare $\sum_i v_i(s_i)$.

The problem becomes computationally challenging when the number of items $m$ is "large", i.e., when the running time of the mechanism is not allowed to be polynomial in $m$ but rather just in $\log m$. There are two variant models in this case, the first assumes that the valuation functions are given as "black boxes" that the algorithm may query[3], and the second assumes that the valuation functions are given in some succinct bidding language. Finding the optimal allocation is essentially a knapsack problem and is computationally hard in both models: in the black-box model it requires exponentially many queries, and in the succinct representation model, it is NP-hard. Just like Knapsack, the optimal social welfare can be approximated arbitrarily well (in both models) and has an FPTAS: approximation ratio of $1+\epsilon$ obtained in time that is polynomial in $n$, $\log m$, and $\epsilon^{-1}$. This FPTAS does not imply any incentive compatible approximation though, and the question boils down to what degree of approximation can be obtained in an incentive compatible way in polynomial time.

Already in Vickrey's seminal paper [32] multi-unit auctions were considered, restricted to the case of downward sloping valuations, i.e., $v_i(k+1) - v_i(k) \leq v_i(k) - v_i(k-1)$ for all $0 < k < m$. For this case the optimal allocation can be found efficiently, as an "equilibrium price" exists, which can be found by binary search (together with the optimal allocation it implies), and attaching the Vickrey payments – the point of his paper – gives incentive compatibility. For general valuations the exact optimum is computationally hard to achieve, so further work considered approximations. The single-dimensional "single minded" case was shown to have a truthful FPTAS [5], improving an earlier 2-approximation [25]. In addition, a PTAS exists for somewhat richer valuations that can be described using certain "bidding languages" (e.g., $k$-minded bidders for a fixed $k$) [14]. The general case was studied in [14] where a truthful 2-approximation was obtained using a maximal-in-range VCG mechanism[4]. It was also shown there that no computationally-efficient maximal in range VCG mechanism can achieve a better approximation ratio.

The main open problem was whether there exist *non-VCG* truthful mechanisms with a better approximation ratio. Some evidence [22, 17] supported the conjecture that that there are no such mechanisms: truthful mechanisms *for two players that always allocate all items* must be affine maximizers[5]. It should be emphasized that the question regards deterministic mechanisms, as a randomized FPTAS was obtained by [12][6].

**Our Results**

Given the evidences and our own intuition, we were surprised to find that there are non-VCG mechanisms that provide arbitrarily good approximation ratios:

**Theorem:** For every $\epsilon > 0$, there exists a truthful $(1+\epsilon)$-approximation mechanism for multi-unit auctions of $m$ items between two players which is not an affine maximizer.

We call these mechanisms *Triage mechanisms* as they split the valuation domain into three sub-domains, depending on the ratios $v(1)/v(m)$ and $v(m-1)/v(m)$. Their payment structure mimics VCG prices with two exceptions: in the "low sub-domain", the price for a single item is decreased to a specific fraction of the value of all items, and in the "high sub-domain", the payments of all non-empty bundles are increased, by the same amount, in a specific linear way. This family of mechanisms is parameterized by three parameters (specifying a weight and the "high thresholds" for both players), with all other parameters uniquely determined by them. We also exhibit two other families of finitely approximating incentive compatible mechanisms, but their approximation factor is worse.

---

[3] The usual query assumed is a "value query", asking for $v_i(k)$ for some $k$, but most lower bounds hold for any queries, as they apply in the communication complexity model.

[4] The situation here mirrors, with different parameters, that of other types of combinatorial auctions where there is a gap between the computationally achievable approximation ratio and the best incentive compatible mechanism known for the multi-parameter case, which is a maximal-in-range VCG mechanism [15, 18, 14].

[5] The driving force of these and similar characterization results is the annulment of the no-externalities condition as everything not won by one player must be given to the other.

[6] This again mirrors the situation in other types of combinatorial auctions where randomized mechanisms are known with better approximation ratios than those obtained by deterministic ones [11, 12].

Our next, and main, result shows that these Triage mechanisms are the only *scalable* incentive compatible mechanisms with a good approximation ratio for the case of two items and two players. Scalability means that the auction's allocation rule does not depend on the "units" in which the valuations are given: multiplying all valuations by the same positive constant does not change the allocation[7]. Triage mechanisms and the other mechanisms mentioned above are all scalable.

**Theorem:** A scalable truthful $c$-approximation mechanism, for $c < 2$, in a multi-unit auction of two items among two bidders must be a Triage mechanism for some choice of parameters.

This is the first characterization of truthfulness in an auction domain or, more generally, in a domain with no externalities. Our novel approach is radically different than previous characterization results (e.g., [30]): we analyze the *payment functions* of the players, rather than the allocation rule directly. The proof is quite involved and reveals the properties of the payment functions (monotonicity, continuity, invertibility, linearity, etc.) gradually, one property after the other. The proof also makes repeated use of the *approximation guarantee* of the mechanism, in contrast to previous results that characterized *all* mechanisms in a certain domain, and were not conditioned on the approximation ratio.

Triage mechanisms are affine maximizers on the "middle sub-domain" and we show that this extends to auctions of an arbitrary number of items among two players.

**Theorem:** A scalable truthful $c$-approximation mechanism, for $c < 2$, in a multi-unit auction of $m > 2$ items among two bidders, must be identical to an affine maximizer with VCG payments on the sub-domain where $v_i(1) = 0$ and $v_i(m-1) = v_i(m-2)$ for every player $i$.

Adopting the point of view of economics, our theorem can be interpreted as follows: Green and Laffont [19] characterize efficient (read: welfare-maximizing) mechanisms and show that VCG is the unique efficient mechanism. We relax the efficiency requirement to "approximate efficiency" and (almost completely) characterize all truthful (scalable) mechanisms in the multi-unit auction domain.

Interestingly, the theorem is not proved by direct characterization, but rather by reducing the characterization problem to the two-item case. We achieve this by introducing a new technical tool that enables us to use our two-item characterization as a black box: *induced mechanisms*. The technique might be of independent interest: it hints that in general characterizing truthful mechanisms may require only the characterization of small instances. The theorem immediately implies computational hardness, a first-of-a-kind result for an auction domain:

**Theorem:** Fix a model of computation in which finding the exact social-welfare maximizing allocation of $m$ items between two players is computationally hard, even with valuations restricted to $v_i(1) = 0$ and $v_i(m-1) = v_i(m-2)$. Then, getting a scalable truthful $c$-approximation, for $c < 2$, of the social welfare in a multi-unit auction of $m$ items among any $n \geq 2$ bidders, is also computationally hard.

This implies an exponential lower bound on communication in the black-box model [14] and implies NP-hardness in the succinct representation model, with, e.g., the bidding language allowing valuations to be specified by boolean circuits [22][8].

Very recently a different approach was introduced for proving the impossibility of polynomial-time truthful mechanisms for combinatorial auctions with submodular bidders that use only value queries [10]. However, we do not know how to apply the technique of [10] to multi-unit auctions. Also note that, unlike [10], the results in this paper are not restricted to a specific type of query. Furthermore, we believe that obtaining characterizations of truthful mechanisms, whenever possible, is of interest regardless of computational considerations.

---

[7]In terms of pure computation, scalability comes for free as one can always scale all inputs by the largest value. We also note that the truthful randomized FPTAS of [12] is scalable.

[8]As expected, the theorem does not imply hardness for, say, single minded bidders, since finding the welfare-maximizing allocation among two single-minded bidders is computationally easy.

The main open problem is to get rid of the scalability assumption which we believe is not really necessary for all our theorems. We note that our reduction to the two-item case from an arbitrary number of items does not require scalability, so the hurdle is really just in characterizing the two-item two-bidder case. The fixed small size would perhaps suggest a direct attack, perhaps even a computer-assisted one, but obviously we were not able to do so.

**Organization**

In Section 2 the setting and basic definitions are given. Triage auctions (and two additional families of auctions) are discussed in Section 3. Section 4 characterizes two-item two-bidder truthful and scalable mechanisms. Finally, Section 5 provides a characterization of mechanisms for any number of items.

## 2 Preliminaries

**The Setting**

In a multi-unit auction there is a set of $m$ identical items, and a set $N = \{1, 2, \ldots, n\}$ of bidders. Each bidder $i$ has a valuation function $v_i : [m] \to \mathbb{R}^+$, which is normalized ($v_i(0) = 0$) and non-decreasing. Denote by $V$ the set of all normalized an non-decreasing valuations. An allocation of the items $\vec{s} = (s_1, \ldots, s_n)$ is a vector of non-negative integers such that $\Sigma_i s_i \leq m$. Denote the set of allocations by $S$. The goal is to find an allocation that maximizes the welfare: $\Sigma_i v_i(s_i)$.

In most of this paper we concentrate in the case where $n = 2$. For convenience, we name the bidders Alice and Bob. We usually denote Alice's valuation by $v$, and Bob's by $u$.

**Truthfulness**

The reader is referred to [26] for the (standard) proofs missing in this subsection. An $n$-bidder mechanism for multi-unit auctions is a pair $(A, p)$ where $A : V^n \to S$ and $p = (p^{(1)}, \cdots, p^{(n)})$, where for each $i$, $p^{(i)} : V^n \to \mathbb{R}$.

**Definition 2.1** *Let $(A, p)$ be a mechanism. $(A, p)$ is* truthful *if for all $i$, all $v_i, v'_i$ and all $v_{-i}$ we have that $v_i(A(v_i, v_{-i})_i) - p^{(i)}(v_i, v_{-i}) \geq v_i(A(v'_i, v_{-i})_i) - p^{(i)}(v'_i, v_{-i})$.*

It is well known that an algorithm (for multi-unit auctions) is truthful if and only if each bidder is presented with a payment for each bundle $t$ that does not depend on bidder $i$'s valuation (i.e., $p^{(i)} : V^{n-1} \to \mathbb{R}$). Denote this payment by $p_t^{(i)}(v_{-i})$. Each bidder is allocated a bundle that maximizes his profit: $v_i(t) - p_t^{(i)}(v_{-i})$ (this is called the "taxation principle" – we will sometimes say that these payments are *induced* by $v_{-i}$). We note that we may assume without loss of generality that for $t > t'$, $p_t^{(1)}(v) \geq p_{t'}^{(1)}(v)$ ("payment monotonicity"): otherwise, we have a mechanism with the same allocation rule by using $p_t^{(1)}(v) = p_{t'}^{(1)}(v)$ and the appropriate tie-breaking between bundles $t$ and $t'$ when $u(t) = u(t')$. The following definition and proposition are standard:

**Definition 2.2** *$A$ is an* affine maximizer *if there exist a set of allocations $\mathcal{R}$, a constant $\alpha_i \geq 0$ for each $i \in N$, and a constant $\beta_{\vec{s}} \in \Re$ for each $\vec{s} \in S$, such that $A(v_1, ..., v_n) \in \arg\max_{\vec{s}=(s_1,...,s_n)\in\mathcal{R}}(\Sigma_i(\alpha_i v_i(s_i)) + \beta_s)$. $A$ is called* welfare maximizer *if $\beta_{\vec{s}} = 0$ for each $\vec{s} \in S$.*

**Proposition 2.3** *Let $A$ be an affine maximizer (in particular, welfare maximizer). There are payments $p$ such that $(A, p)$ is a truthful mechanism.*

Notice that when $A$ is a two-bidder welfare maximizer, the payments are as follows: there is a constant $w > 0$ such that for each $t$, a valuation $v$ of Alice and a valuation $u$ of Bob, $p_t^{(2)}(v) = w(v(m) - v(m-t))$ and $p_t^{(1)}(u) = (u(m) - u(m-t))/w$. We sometimes use a table notation to denote a 2-item instance. This notation is illustrated below for the 2-bidder welfare maximizer case (notice that each bidder's pavements depend only on the valuation of the other bidder):

| Number of items | Alice's value | Alice's payment | Bob's value | Bob's payment |
|---|---|---|---|---|
| One | $v(1)$ | $(u(2) - u(1))/w$ | $u(1)$ | $w(v(2) - v(1))$ |
| Two | $v(2)$ | $u(2)/w$ | $u(2)$ | $w \cdot v(2)$ |

**Scalability**

This paper considers two definitions of scalability.

**Definition 2.4** *An auction is* allocation scalable *if multiplying the valuations of all bidders by the same positive factor does not change the allocation.*

**Definition 2.5** *An auction is* payment scalable *if for each bidder $i$, valuations of the other bidders $v_{-i}$, and $\alpha > 0$, $\alpha \cdot p^{(i)}(v_{-i}) = p^{(i)}(\alpha \cdot v_{-i})$.*

In the appendix (Proposition A.1) we show that every allocation scalable mechanism is also payment scalable, and thus in this paper we use the term *scalable* to denote the less restrictive notion of scalability – payment scalability.

## 3 The Triage Auction

We present three families of truthful mechanisms for multi-unit auctions that provide a bounded approximation ratio for multi-unit auctions with two bidders. Each of the families contain mechanisms that are not affine maximizers. The first family, the Triage auction, includes mechanism that guarantee an approximation ratio of $1 + \epsilon$, and the next sections show that triage auctions are the only two-item truthful and scalable mechanisms that provide an approximation ratio better than 2. The other two families – shifted welfare maximizers and fractions auctions – provide an approximation of almost 2. We postpone their description to the appendix. To the best of our knowledge all previously known finitely-approximating mechanisms are either affine maximizers or are essentially single-parameter mechanisms (i.e., each bidder either receives all items, or no items at all).

We describe the mechanisms by specifying the payment functions of the bidders (recall that each function depends only on the other bidder's valuation). Truthfulness is obvious since each bidder is allocated a bundle that maximizes his profit, and we are left only with proving feasibility and analyzing the approximation ratio. The proof of the theorem below appears in the appendix.

**Definition 3.1** *The Triage auction is parameterized by three parameters, $w, \theta_A, \theta_B$, for $w > 0$, $0 \leq \theta_A, \theta_B \leq 1$, and $\theta_A \geq 1 - \theta_B$. The payment functions are:*

- $p_m^{(2)}(v) = wv(m)$ *if* $v(1) < \theta_A v(m)$, *and* $p_m^{(2)}(v) = \frac{wv(1)}{\theta_A}$ *otherwise.*

- *For* $2 \leq k \leq m - 1$, $p_k^{(2)}(v) = p_m^{(2)}(v) - w \cdot v(m - k)$.

- $p_1^{(2)}(v) = p_m^{(2)}(v) - wv(m-1)$ *if* $v(m-1) > (1 - \theta_B)v(m)$, *and* $p_1^{(2)}(v) = p_m^{(2)}(v) - w(1 - \theta_B)v(m)$ *otherwise (notice that in the latter case we have in fact $p_m^{(2)}(v) = wv(m)$).*

*and*

- $p_m^{(1)}(u) = w^{-1}u(m)$ *if* $u(1) < \theta_B u(m)$, *and* $p_m^{(1)}(u) = \frac{w^{-1}u(1)}{\theta_B}$ *otherwise.*

- *For* $2 \leq k \leq m-1$, $p_k^{(1)}(v) = p_m^{(1)}(u) - w^{-1} \cdot u(m-k)$.

- $p_1^{(1)}(u) = p_m^{(1)}(u) - w^{-1}u(m-1)$ *if* $u(m-1) > (1-\theta_A)u(m)$, *and* $p_1^{(1)}(v) = p_m^{(1)}(u) - w^{-1}(1-\theta_A)u(m)$ *otherwise (again, in the latter case* $p_m^{(1)}(u) = wu(m)$).

**Theorem 3.2** *The* $(w, \theta_A, \theta_B)$-*Triage auction is feasible. The* $(1, \theta_A, \theta_B)$-*Triage auction provides an approximation ratio of* $\max(\frac{1}{\theta_A}, \frac{1}{\theta_B})$.

We remark that when $w = \theta_A = \theta_B = 1$ we get the VCG mechanism.

## 4 Characterization of Scalable 2-Item Auctions

This section is devoted to proving the following characterization result:

**Theorem 4.1 (two-item characterization)** *The only feasible, scalable and truthful auctions with an approximation ratio strictly better than 2 for two identical goods and two bidders are triage auctions for some* $(w, \theta_A, \theta_B)$.

We now provide a brief road map to the proof of the theorem. Very differently from Roberts' theorem proof, we analyze the *payment functions* of the bidders (rather then the allocation rule) and show that the payment functions are identical to the payment functions of some triage auction. Recall that in the two-item case, the payment functions of a triage auction are defined using three different regions that correspond to the ratio between the value for two items and the value for one item: high, mid, and low. The proof of the theorem is quite involved and for readability we divide it into subsections that roughly correspond to these regions.

Subsection 4.1 gives an alternative definition of the triage auction, for the special case where $m = 2$, that is easier for us to work with. Subsection 4.2 characterizes the payment function for two items. The results of subsection 4.1 hold for any scalable mechanism with a bounded approximation ratio, not just ones with an approximation ratio better than 2. The next subsections are devoted to characterizing the payment functions for one item. Subsection C.2 defines and "separates" the high-range from the mid and low ranges: it shows that, roughly speaking, a valuation that is not in the high range induces payment (for one item) that is also not in the high range. Subsection C.3 proves some basic properties, like continuity, of the payment function in the low and mid range. The central part of the proof is Subsection C.4 which shows that the payment functions in the mid range are equivalent to the payment functions of weighted VCG. We conclude the proof with Subsection C.5 that handles the value of the transition points between the high and mid range, and Subsection C.6 that handles the high range.

Due to lack of space we defer the last sections of the proof to the appendix, and keep here only the simpler Subsection 4.2.

### 4.1 An Alternative Description of the Triage Auction with Two Items

**Definition 4.2** *Let* $p, q : [0,1] \to \Re^+$ *and* $f, g : [0,1] \to [0,1]$ *be real valued functions. The scalable mechanism based on them is given by the following table.*

| Number of items | Alice's value | Alice's payment | Bob's value | Bob's payment |
|---|---|---|---|---|
| One | $rv$ | $g(s) \cdot q(s) \cdot u$ | $su$ | $f(r) \cdot p(r) \cdot v$ |
| Two | $v$ | $q(s) \cdot u$ | $u$ | $p(r) \cdot v$ |

**Proposition 4.3** *For any $p, q : [0, 1] \to \Re^+$ and $f, g : [0, 1] \to [0, 1]$, the scalable mechanism based on them is scalable and truthful (but may be infeasible and allocate more than 2 items). Any truthful scalable mechanism (even a non-feasible one as long as it allocates at most two items to any bidder) is equivalent to one based on some functions.*

The proof of the proposition can be found in the appendix. We now give an alternative (equivalent) definition of the triage auction, for the $m = 2$ case.

**Definition 4.4** *The $(w, \theta_A, \theta_B)$-triage auction for $w > 0$ and $0 \leq \theta_A, \theta_B \leq 1$, $\theta_A \geq 1 - \theta_B$, is the scalable mechanism based on:*

- *For $r \leq 1 - \theta_B$: $f(r) = \theta_B$, and $p(r) = w$.*
- *For $1 - \theta_B \leq r \leq \theta_A$: $f(r) = 1 - r$, and $p(r) = w$.*
- *For $r \geq \theta_A$: $f(r) = 1 - \theta_A$, and $p(r) = wr/\theta_A$.*

*and*

- *For $s \leq 1 - \theta_A$: $g(s) = \theta_A$, and $q(s) = w^{-1}$.*
- *For $1 - \theta_A \leq s \leq \theta_B$: $g(s) = 1 - s$, and $q(s) = w^{-1}$.*
- *For $s \geq \theta_B$: $g(s) = 1 - \theta_B$, and $q(s) = w^{-1}s/\theta_B$.*

## 4.2 Characterizing the Payment for Two Items

The results in this section hold for any scalable and truthful mechanism with a bounded approximation ratio. We usually prove the theorem only for the function $p$. The proof for $q$ is symmetric. We remind the reader that the following sections of the proof are in the appendix.

**Lemma 4.5** *The function $p$ is monotone non-decreasing.*

**Proof:** Assume towards contradiction that for some $r' > r$ we have $p(r') < u < u' < p(r)$. Since $u > p(r')$, on inputs $(r', 1)$ and $(0, u)$ Bob must win both items, so Alice cannot win anything. Notice that $(r, 1)$ wins nothing against $(0, u'(1 + \epsilon))$ (by payment scalability, $(0, u'(1 + \epsilon))$ induces bigger payments than $(0, u')$, and Alice did not win any items with the bigger valuation $(r, 1)$), but also Bob does not win both items since $u'(1 + \epsilon) < p(r)$ for small enough $\epsilon > 0$, so the total welfare achieved is 0 contradicting finite approximation. □

**Lemma 4.6 (weighting)** $p(0) \cdot q(0) = 1$.

**Proof:** Consider the following input:

| Number of items | Alice's value | Alice's payment | Bob's value | Bob's payment |
|---|---|---|---|---|
| One | 0 | ? | 0 | ? |
| Two | $v$ | $uq(0)$ | $u$ | $vp(0)$ |

The only allocations that give finite approximation ratio on inputs of the form $(0, v)$ and $(0, u)$ are those that give both items to one of the bidders. If $u < vp(0)$ then Bob does not win two items; whereas if $u > vp(0)$ then he wins both items, and dually for Alice. So we get a contradiction to feasibility if $u > vp(0)$ and $v > uq(0)$, i.e., if $p(0)q(0) < 1$. On the other hand, if $u < vp(0)$ and $v < uq(0)$, i.e., if $p(0)q(0) > 1$, then we get a total welfare of 0, contradicting finite approximation ratio. □

At this point we are ready to give a more precise definition of the payment function. We start with the low range, i.e., when $r < g(0)$. In particular we show that the function is constant in this range.

**Lemma 4.7 (low range)** *If $r < g(0)$ then $p(r) = p(0)$.*

**Proof:** Assume that $p(r) \neq p(0)$ then, using monotonicity, let $p(r) > u' > u > p(0)$. On input $(0,1)$ and $(0,u)$ Bob gets both items (since $u > p(0)$) and so Alice must get none. On inputs $(r,1)$ and $(0,u')$ Alice gets at most 1 item (since, by the scalability of the payments, the payment induced by Bob for two items has increased), but since $u' < p(r)$ Bob does not get two items, and so for finite approximation, Alice must get an item, so $r \geq u'g(0)q(0) \geq p(0)q(0)g(0) = g(0)$. □

The following claim would be helpful in analyzing the payment function in the high range:

**Claim 4.8** $p(r) \geq r/(g(0)q(0))$.

**Proof:** Let $u > p(r)$, then on input $(r,1)$ and $(0,u)$ Bob gets both items. Alice's payment for a single item is $ug(0)q(0)$ which for feasibility must be at least $r$. Since this holds for all $u > p(r)$ we get that $r \leq p(r)g(0)q(0)$ as required. □

For the high range ($r > g(0)$) we show that the payment grows in a specific linear way:

**Lemma 4.9 (high range)** *If $r > g(0)$ then $p(r) = r/(g(0)q(0))$.*

**Proof:** We will prove the contra-positive which by the previous claim assumes $p(r) > u > r/(g(0)q(0))$. Consider the following input:

| Number of items | Alice's value | Alice's payment | Bob's value | Bob's payment |
|---|---|---|---|---|
| One | $r$ | $uq(0)g(0)$ | 0 | ? |
| Two | 1 | $uq(0)$ | $u$ | $p(r)$ |

In this case Bob cannot win both items so he gets a value of 0. By the choice of $u$, Alice's payment for two items is $uq(0) > r/g(0)$ is greater than 1, thus she cannot win two items. Thus for finite approximation she must win one item and thus $r \geq uq(0)g(0)$ and since this is true for every $u < p(r)$, we have $r \geq p(r)g(0)q(0)$, contradiction. □

At this point we have completed the required characterization of $p$ and $q$.

**Definition 4.10** *Let $w = p(0)$, $\theta_A = g(0)$, and $\theta_B = f(0)$.*

**Lemma 4.11 (summary of subsection)** *For $r \leq \theta_A$ we have that $p(r) = w$ and for $r \geq \theta_A$ we have $p(r) = wr/\theta_A$. Similarly, for $s \leq \theta_B$ we have that $q(s) = w^{-1}$ and for $s \geq \theta_B$ we have $q(s) = w^{-1}s/\theta_B$.*

**Proof:** The low range lemma states the required fact for $r < \theta_A$. The high range lemma states the required fact for $r \geq \theta_A$, taking into account the inverse lemma, $p(0)q(0) = 1$, the same holds for $q$, replacing $w$ with $w^{-1}$, again relying on $p(0)q(0) = 1$. For $r = \theta_A$ we observe that $p(r) = w$ since $p$ is a monotone function and approaches $w$ above and below $w$. □

## 5 Characterizing Mechanisms for any Number of Items

We showed that every two-item scalable mechanism that provides an approximation ratio better than 2 is a triage auction. This section gives an almost complete characterization for truthful and scalable mechanisms that guarantee an approximation ratio better than 2 for any number of items. In particular this section's characterization implies that truthful and scalable mechanisms for multi-unit auctions cannot guarantee an approximation ratio better than 2 in polynomial-time.

The two-item characterization is used as a black box to characterize mechanisms for more items. Importantly, the scalability assumption is not used in this section. In other words, proving that triage auctions are the only truthful mechanisms (scalable or not) that provide an approximation ratio better than 2 in multi-unit auctions with only two items, would immediately imply our characterization result for any number of items, and in particular would imply an unconditional lower bound on the power of all polynomial time truthful mechanisms. All missing proofs appear in the appendix.

## 5.1 Induced Mechanisms: Definition and Basic Properties

Our main working horses will be induced mechanisms. Induced mechanisms allow us to define a two-item mechanism given an $m$-item mechanism. By leveraging our two-item characterization, we show that the induced two-item mechanisms are triage auctions. We then study the relationship between all induced mechanisms and prove that many of them must be welfare maximizers. We show that this implies that the $m$-item mechanism we started with must have a very specific form, as needed.

**Definition 5.1** *Let $l_1, h_1$ be such that $1 \leq l_1 < h_1 \leq m$. The $(l_1, h_1)$-extension of a two-item valuation $v$, denoted $v^{l_1,h_1}$, is defined as follows: for every $k < l_1$, $v^{l_1,h_1}(k) = 0$. For every $h_1 > k \geq l_1$, $v^{l_1,h_1}(k) = v(1)$. For every $k \geq h_1$, $v^{l_1,h_1}(k) = v(2)$.*

**Definition 5.2 (Induced Mechanism)** *Let $A$ be a mechanism for multi-unit auctions with $m$ items and $2$ bidders. Let $l_1, h_1, l_2, h_2$ be positive integers with the following constraints: $l_1 < h_1 \leq m$, $l_2 < h_2 \leq m$, $l_1 + l_2 \leq m$, $l_1 + h_2 > m$ and $l_2 + h_1 > m$. Define the induced mechanism $A^{l_1,h_1,l_2,h_2}$ for $2$ items as follows: given two valuations $v$ and $u$ run $A$ with the $(l_1, h_1)$-extended valuation $v^{l_1,h_1}$ and the $(l_2, h_2)$-extended valuation $u^{(l_2,h_2)}$. Let $(a_1, a_2)$ be the output allocation of $A$ and $(p_1, p_2)$ be the payments the bidders are charged in $A$. If $a_1 < l_1$ then let $a'_1 = 0$, if $l_1 \leq a_1 < h_1$ then let $a'_1 = 1$, otherwise let $a_1 = 2$. If $a_2 < l_2$ then let $a'_2 = 0$, if $l_2 \leq a_2 < h_2$ then let $a'_2 = 1$, otherwise let $a_2 = 2$. The output of $A^{l_1,h_1,l_2,h_2}$ on $v$ and $u$ is $(a'_1, a'_2)$. Alice's payment is $p_1$ and Bob's payment is $p_2$.*

**Proposition 5.3** *Let $A$ be a scalable mechanism for multi-unit auctions with $m$ items and $2$ bidders that provides an approximation ratio of $\alpha$. Let $A^{l_1,h_1,l_2,h_2}$ be an induced mechanism. $A^{l_1,h_1,l_2,h_2}$ is feasible, truthful, scalable, and provides an approximation ratio of $\alpha$.*

In this section we denote the $p^{(2)}$ function of $A$ (the payments induced by Alice) by $f$ and by $f^{l_1,h_1,l_2,h_2}$ the $p^{(2)}$ function of the induced mechanism $A^{l_1,h_1,l_2,h_2}$. We denote by $g$ the $p^{(1)}$ function of $A$ (the payments induced by Bob) and by $g^{l_1,h_1,l_2,h_2}$ the $p^{(1)}$ function of the induced mechanism $A^{l_1,h_1,l_2,h_2}$. As a corollary of Proposition 5.4 we get the following relationship between the payment functions of $A$ and its induced mechanisms.

**Corollary 5.4** *Let $v$ be a two-item valuation and let $v^{(l_1,h_1)}$ be its $(l_1, h_1)$-extension. Let $l_2, h_2$ be such that $A^{l_1,h_1,l_2,h_2}$ is an induced mechanism. $f_{l_2}(v^{h_1,l_1}) = f_1^{l_1,h_1,l_2,h_2}(v)$ and $f_{h_2}(v^{h_1,l_1}) = f_2^{l_1,h_1,l_2,h_2}(v)$.*

*Symmetrically, let $u$ be a two-item valuation and let $u^{(l_2,h_2)}$ be its $(l_2, h_2)$-extension. Let $l_1, h_1$ be such that $A^{l_1,h_1,l_2,h_2}$ is an induced mechanism. $g_{l_1}(u^{h_2,l_2}) = g_1^{l_1,h_1,l_2,h_2}(u)$ and $g_{h_1}(u^{h_2,l_2}) = g_2^{l_1,h_1,l_2,h_2}(u)$.*

## 5.2 Relations between Induced Mechanisms

Let $A$ be a scalable and truthful mechanism for multi-unit auctions for $m$ items with an approximation ratio better than 2. By our characterization and the discussion above above we have that all induced mechanisms of $A$ are triage auctions. Denote the parameters of the triage mechanism $A^{l_1,h_1,l_2,h_2}$ by $\theta_A^{l_1,h_1,l_2,h_2}, \theta_B^{l_1,h_1,l_2,h_2}, w^{l_1,h_1,l_2,h_2}$. The point of this subsection is to study the relations between the parameters of the induced triage mechanisms of $A$.

**Claim 5.5** *Let $A$ be a truthful and scalable mechanism for multi unit auctions with $m$ items that provides an approximation ratio better than 2. Let $A^{l_1,h_1,l_2,h_2}$ and $A^{l_1,h_1,l_2,h'_2}$ be two induced mechanisms of $A$. Then, $w^{l_1,h_1,l_2,h_2} = w^{l_1,h_1,l_2,h'_2}$, $\theta_A^{l_1,h_1,l_2,h_2} = \theta_A^{l_1,h_1,l_2,h'_2}$, and $\theta_B^{l_1,h_1,l_2,h_2} = \theta_B^{l_1,h_1,l_2,h'_2}$. Symmetrically, let $A^{l_1,h_1,l_2,h_2}$ and $A^{l_1,h'_1,l_2,h_2}$ be two induced mechanisms of $A$. Then, $w^{l_1,h_1,l_2,h_2} = w^{l_1,h'_1,l_2,h_2}$, $\theta_A^{l_1,h_1,l_2,h_2} = \theta_A^{l_1,h'_1,l_2,h_2}$, and $\theta_B^{l_1,h_1,l_2,h_2} = \theta_B^{l_1,h'_1,l_2,h_2}$.*

**Claim 5.6** *Let $A$ be a truthful and scalable mechanism for multi unit auctions with $m$ items that provides an approximation ratio better than 2. Let $A^{l_1,h_1,l_2,h_2}$ and $A^{l_1,h_1,l'_2,h_2}$ be two induced mechanisms of $A$. Then, $\theta_A^{l_1,h_1,l_2,h_2} = \theta_A^{l_1,h_1,l'_2,h_2}$ and $w^{l_1,h_1,l_2,h_2} = w^{l_1,h_1,l'_2,h_2}$. Symmetrically, let $A^{l_1,h_1,l_2,h_2}$ and $A^{l'_1,h_1,l_2,h_2}$ be two induced mechanisms of $A$. Then, $\theta_B^{l_1,h_1,l_2,h_2} = \theta_B^{l'_1,h_1,l_2,h_2}$ and $w^{l_1,h_1,l_2,h_2} = w^{l'_1,h_1,l_2,h_2}$.*

We now use the claims to prove that all induced mechanisms share the same $w$. Thus, after proving it we may denote the $w^{l_1,h_1,l_2,h_2}$ parameter of every induced mechanism $A^{l_1,h_1,l_2,h_2}$ by (the same) $w$.

**Lemma 5.7** *Let $A$ be a truthful and scalable mechanism for multi unit auctions with $m$ items that provides an approximation ratio better than 2. Let $A^{l_1,h_1,l_2,h_2}$ and $A^{l'_1,h'_1,l'_2,h'_2}$ be two induced mechanisms of $A$. Then, $w^{l_1,h_1,l_2,h_2} = w^{l'_1,h'_1,l'_2,h'_2}$.*

## 5.3 Some Induced Mechanisms are Welfare Maximizers

The heart of this section is here. We show that "many" of the induced triage auctions take the simple form of welfare maximizers. The crux is that the payment function for some items is simultaneously the payment function of one item for one induced mechanism, and the payment function for two items for another. Simple algebra then gives us that some $\theta$'s must equal to 1. We are then able to specify the payment functions of "simple" valuations.

**Lemma 5.8** *Let $A$ be a truthful and scalable mechanism for multi unit auctions with $m$ items with an approximation ratio better than 2. Let $A^{l_1,m,l_2,m}$ where $l_1, l_2 \geq 2$. Then, $\theta_A^{l_1,m,l_2,m} = \theta_B^{l_1,m,l_2,m} = 1$.*

**Definition 5.9** *A valuation $v$ is $l$-simple if there exists some $0 < l < m$ such that for every $k < l$, $v(k) = 0$, and for every $l \leq k < m$ we have that $v(k) = v(l)$.*

**Corollary 5.10** *Let $l \geq 2$. For every $l$-simple valuation $v$ we have that $f_m(v) = wv(m)$ and for all $1 < t < m - 1$ such that $l + t \leq m$ we have that $f_t(v) = w(v(m) - v(m - t))$. Similarly, for every $l$-simple valuation $u$ we have that $f_m(u) = u(m)/w$ and for all $1 < t < m - 1$ such that $l + t \leq m$ we have that $g_t(u) = (u(m) - u(m - t))/w$.*

**Proof:** For the first statement, observe that $A^{l,m,t,m}$ is an induced mechanism for every $t$ such that $l + t \leq m$ and apply Lemma 5.8. The proof of the second statement is similar. □

## 5.4 Concluding the Characterization

We are now ready to obtain our final characterization. We give an almost complete characterization of the payment function for valuations where the value for one item is 0. Lemma 5.11 provides the payment function for $m$ items, and using it Lemma 5.12 the payment functions for smaller bundles.

**Lemma 5.11** *For each $v$ where $v(1) = 0$, $f_m(v) = wv(m)$. Symmetrically, for each $u$ where $u(1) = 0$, $g_m(u) = \frac{u(m)}{w}$.*

**Lemma 5.12** *Let $v$ be a valuation with $v(1) = 0$. For every $k \neq 1, m - 1$ we have that $f_k(v) = w(v(m) - v(m - k))$. Similarly, for every valuation $u$ with $u(1) = 0$ we have, for every $k \neq 1, m - 1$ that $g_k(u) = w^{-1}(u(m) - u(m - k))$.*

Our final characterization result is:

**Definition 5.13** *A valuation $v$ is* degenerate *if $v(1) = 0$ and $v(m-1) = v(m-2)$.*

**Theorem 5.14 (Characterization of mechanisms for any number of items)** *Let $A$ be a truthful and scalable mechanism for $m > 2$ items and two bidders that provides an approximation ratio better than $2$. There exists a constant $w > 0$ such that on all inputs $(v, u)$ where $v$ and $u$ are degenerate, $A$ outputs a solution with value $\max_k(v(k) + wu(m-k))$.*

**Acknowledgements**

The first author wishes to thank Itai Ashlagi and Ron Lavi for several stimulating early discussions. The fractions auction of Section 3 was obtained during these discussions. We thank Bobby Kleinberg for helpful discussions. We thank Sigal Oren for numerous very helpful comments and ideas throughout the work on this paper.

# A  Appendix for Section 2

**Proposition A.1** *Let A be an allocation scalable mechanism. Then, A is also payment scalable.*

We prove the lemma for the case of $n = 2$ but the proof easily extends to $n > 2$ bidders. Fix a valuation $u$ of Bob. Let $B_t(u)$ be the set of all valuations $v$ that assign Alice $t$ items in input $(v, u)$. Formally, $B_t(u) = \{v | A_1(v, u) = t\}$. We say that $t$ is *in the range* of $u$ if $B_t \neq \emptyset$.

We claim that $t$ is in the range of $u$ if and only if $t$ is in the range of $\alpha \cdot u$. To see that, consider $t$ that is in the range of $u$. We have that $t$ is also in the range of $\alpha \cdot u$ since $A(v, u) = A(\alpha \cdot v, \alpha \cdot u)$. The 'only if' direction is symmetric.

We now show that for every $t, t'$ in the range of $u$ and $\alpha > 0$, $\alpha(p_t^{(i)}(u) - p_{t'}^{(i)}(u)) = p_t^{(i)}(\alpha \cdot u) - p_{t'}^{(i)}(\alpha \cdot u)$ (for bundles not in the range we set the payment to be equal to the payment of the next bigger bundle that is in the range, and use the appropriate tie-breaking rule). Fix some $t$ in the range of $u$ such that there exists $v \in B_t(u)$ that is on the border of $B_t(u)$ (in the usual topological sense). If there is no such point, Alice is always assigned $t$ and we are done by letting $p_i^{(i)}(\alpha \cdot u) = 0$ for every $\alpha \neq 0$. Assume otherwise. There exists at least one $t' \neq t$ which is in every $\epsilon$-neighborhood of $v$ and in $B_{t'}$, since $v$ is on the border. Thus we have $p_t^{(i)}(u) - p_{t'}^{(i)}(u) = v(t) - v(t')$. From scalability, we have that $\alpha \cdot v$ is on the border of $B_t(u)$ with $t'$ playing the same role. We have that $p_t^{(i)}(\alpha \cdot u) - p_{\alpha \cdot t'}^{(i)}(u) = \alpha(v(t) - v(t')) = \alpha(p_t^{(i)}(u) - p_{t'}^{(i)}(u))$.

We continue similarly. Fix $t'' \neq t, t''$ where there exists $v \in B_{t''}(u)$, and $v$ is on the border of $B_t \cup B_{t'}$. Thus, in every $\epsilon$-neighborhood of $v$ there exists a valuation $v'$ for which $v' \in B_t$ or $v' \in B_{t''}$. Without loss of generality assume that $u' \in B_t$ (otherwise, switch the roles of $t$ and $t'$). By using scalability similarly to the arguments above, we get that $v(t) - v(t'') = p_t^{(i)}(u) - p_{t''}^{(i)}(u)$, and consequently we also have $\alpha(v(t) - v(t'')) = p_t^{(i)}(\alpha \cdot u) - p_{t''}^{(i)}(\alpha \cdot u) = \alpha(p_t^{(i)}(u) - p_{t'}^{(i)}(u))$. The proof continues similarly until all bundles in the range are considered.

# B  Appendix for Section 3

Fixing the other bidder's valuation, we say that a bundle of $s$ items is in the *winning set* of a bidder, if this bundle maximizes his profit. We assume that the algorithm chooses an allocation $(s, t)$ with the maximal value such that $s$ is an the winning set of Alice and $t$ is in the winning set of Bob.

## B.1  Proof of Theorem 3.2

We will use the following claim several times:

**Claim B.1** *For triage auctions with $w = 1$, for each optimal allocation $(k, m - k)$, Alice's winning set contains at least one of the following bundles: $k$ items, one item, or no items. Similarly, Bob's winning set contains at least one of the following bundles: $m - k$ items, one item, or no items.*

**Proof:** We will show that the equation $v(k) - p_k^{(2)}(u) \geq v(t) - p_t^{(2)}(u)$ holds for $t \neq 0, 1$. The equation implies that Alice prefers $k$ items over $t$ items, thus if $t$ is in the winning set so does $k$, as needed. To see that, observe that for each $t \neq 1, 0$ we have that $p_k^{(2)}(u) - p_t^{(2)}(u) \geq u(m - k) - u(m - t)$. Since $(k, m - k)$ is an optimal allocation, we also have that $v(k) + u(m - k) \geq v(t) + u(m - t)$. Together we have that $v(k) - p_k^{(2)}(u) \geq v(t) - p_t^{(2)}(u)$, for $t \neq 0, 1$.  □

**Lemma B.2** *The $(w, \theta_A, \theta_B)$-Triage auction is feasible.*

**Proof:** We first note that without loss of generality we may assume that $w = 1$ (since multiplying one bidder's payments by $w$ and dividing the other's by the same $w$ maintains feasibility).

Suppose first that $(k, m - k)$ is an optimal allocation where $k \neq 0, m$. We claim that in this case the mechanism is feasible: by Claim B.1, Alice's winning set contains at least one of the following bundles: 0, 1, or $k$ items. Similarly, Bob's winning set contains at least one of the following bundles: 0, 1, $m - k$. Thus there is a feasible allocation $(s, t)$ such that $s$ is in the winning set of Alice and $t$ is in the winning set of Bob.

Thus from now on we assume that in all optimal allocations at least one bidder is assigned the empty bundle. Suppose that $(m, 0)$ is an optimal allocation (the case where $(0, m)$ is an optimal allocation is similar). Bob's winning set contains either the empty bundle of the bundle of 1 item (since the optimality of $(m, 0)$ implies that $v(m) \geq u(m)$, and if the bundle of $m$ items is in the winning set, so is the empty bundle). Thus, we only have to show that if Bob's winning set contains only the bundle of 1 item, then Alice's winning set contains also bundles that have less than $m$ items.

Bob's winning set will contain only the bundle of 1 item implies that $u(1) > p_1^{(2)}(v)$. The definition of triage auction implies that this means that $u(1) > v(m) - v(m - 1)$ or $u(1) > v(m)\theta_B$, depending on the ratio between $v(m - 1)$ and $v(m)$. The first case cannot happen since it implies that $u(1) + v(m - 1) > v(m)$, which is false since we assumed that $(m, 0)$ is an optimal allocation. In the second case, $p_m^{(1)} = \frac{u(1)}{\theta_B}$, since $u(1) > v(m)\theta_B \geq u(m)\theta_B$. Therefore, the bundle of $m$ items is not in Alice's winning set since $v(m) < \frac{u(1)}{\theta_B}$. □

**Lemma B.3** *The $(1, \theta_A, \theta_B)$-Triage auction provides an approximation ratio of $\max(\frac{1}{\theta_A}, \frac{1}{\theta_B})$.*

**Proof:** Let $(k, m - k)$ be an optimal allocation. In case Alice has the bundle of $k$ items in her winning set and Bob has the bundle of $m - k$ items in his winning set then the approximation ratio is 1. By Claim B.1, the only other cases to consider are when at least one of the bidders (without loss of generality Alice) does not has these bundles (but instead has either the empty bundle or the bundle of one item).

The first case we consider is when Alice's winning set does not contain the bundle of $k$ items, but contains the empty bundle (in particular, we have that $k \neq 0$). Since the empty bundle has a zero profit, it means that the profit from the bundle of $k$ items is negative: $v(k) < p_k^1(u)$. By the definition of the payment function we either have that $p_k^1(u) \geq u(m) - u(m - k)$ or that $p_k^1(u) = \frac{u(1)}{\theta_B} - u(m - k)$.

The first option does not occur since otherwise $v(k) < u(m) - u(m - k)$. In other words, the social welfare of the allocation $(0, m)$ is bigger than the social welfare of the allocation $(k, m - k)$, a contradiction to the optimality of the latter. Thus, the second option occurs and in particular:

$$v(m) < \frac{u(1)}{\theta_B} \tag{1}$$

Since Alice's profit for the bundle of $k$ items is negative, (i.e., $v(k) < \frac{u(1)}{\theta_B} - u(m - k)$), we have, in particular, that $v(m) < \frac{u(1)}{\theta_B}$. Therefore, to prove that an approximation of $\frac{1}{\theta_B}$, it suffices to show that that Bob has at least one non-empty bundle in his winning set. Suppose not, i.e.: $u(1) < p_1^{(2)}(v) \leq \theta_B v(m)$ (the last inequality is from the definition of the payment function – the payment for one item is always at most $\theta_B v(m)$). But then, $\frac{u(1)}{\theta_B} \leq v(m)$, a contradiction to (1).

We are left with considering the case where the only bundle in Alice's winning set is the bundle of one item. We start by showing that $u(m - 1) \leq (1 - \theta_A)u(m)$. Suppose for contradiction that $u(m - 1) > (1 - \theta_A)u(m)$. Since the bundle of one item maximizes the profit: $v(1) - p_1^{(1)}(u) > v(k) - p_k^{(1)}(u)$. Using the definition of the payment function: $v(1) + u(m - 1) > v(k) + u(m - k)$. In other words, the allocation $(k, m - k)$ is not optimal, a contradiction. Thus we have established that

$$u(m - 1) \leq (1 - \theta_A)u(m) \tag{2}$$

Recall that the bundle of one item is profitable for Alice. Using (2) and the definition of the payment function, we claim that:
$$v(1) \geq \theta_A u(m) \tag{3}$$

Since the bundle of one item maximizes the profit, using the definition of the payment function: $v(1) + u(m-1) > v(k) + u(m-k)$. Using (2) we have that:
$$v(1) + (1-\theta_A)(m) > v(k) + u(m-k) \tag{4}$$

Thus, the approximation ratio is no worse than $\frac{v(k)+u(m-k)}{v(1)} \leq \frac{v(1)+(1-\theta_A)u(m)}{v(1)} \leq 1 + \frac{(1-\theta_A)u(m)}{\theta_A u(m)} \leq \frac{1}{\theta_A}$ (where the leftmost inequality is due to (4), and the middle one is due to (3)). □

## B.2 Shifted Welfare Maximizers

**Definition B.4** *An $\alpha$-shifted welfare maximizer is the following two-bidder auction: Bob's payment for $t$ items, $t \neq 0, m$ is $(1+\alpha)v(m) - v(m-t)$, for $t = 0$ items is $0$, and for $t = m$ items is $v(m)$. Alice's payment for $t$ items, $t \neq 0, m$ is $(1+\alpha)u(m) - u(m-t)$, for $t = 0$ items is $0$, and for $t = m$ items is $u(m)$.*

In all proofs we assume without loss of generality that $v(m) \geq u(m)$.

**Lemma B.5** *For any $\alpha > 0$ the $\alpha$-shifted welfare maximizer is feasible.*

**Proof:** Observe that this implies that if the bundle of $m$ items is in Bob's winning set, so does the empty bundle. Suppose that Alice has the bundle of $k$ items in her winning set. We show that Bob has a bundle of size at most $m - k$ items in his winning set, hence the mechanism is feasible. If Alice has the bundle of $k$ items in her winning set, we have that for every $t \neq 0, m$:

$$v(k) - ((1+\alpha)u(m) - u(m-k)) \geq v(t) - ((1+\alpha)u(m) - u(m-t))$$

$$v(k) + u(m-k)) \geq v(t) + u(m-t))$$

Thus we have that Bob (weakly) prefers $m - k$ items over $m - t$ items, for every $t \neq 0, m$:

$$u(m-k) - ((1+\alpha)v(m) - v(k)) \geq u(t) - ((1+\alpha)v(m) - v(t))$$

Note that this concludes the proof for this case: we have already argued that if Bob has the bundle of $m$ items in his winning set he also have the empty bundle, and if Bob is assigned $0$ items feasibility still holds.

We now handle the case where Alice is allocated $m$ items. We have that for every $t \neq 0, m$:

$$v(m) - u(m) \geq v(t) - ((1+\alpha)u(m) - u(m-t))$$

$$v(m) \geq v(t) + u(m-t) - \alpha u(m)$$

$$(1+\alpha)v(m) \geq v(t) + u(m-t)$$

Therefore, the profit of Bob from taking $m - t \neq m, 0$ items is at most $0$:

$$u(m-t) - ((1+\alpha)v(m) - v(t)) \leq 0$$

The proof is now concluded, since if $m$ is in Bob's winning set, so does the empty bundle. □

**Lemma B.6** *For any $1 \geq \alpha > 0$ the $\alpha$-shifted welfare maximizer provides an approximation ratio of $1 + \frac{1}{\alpha}$.*

**Proof:** Observe that when Alice is allocated $m$ items then we have a 2-approximation. Thus assume that Alice is allocated some bundle of $k < m$ items. Suppose that $k = 0$. In this case Alice also has the bundle of $m$ items in her winning set. Notice that Bob has the empty bundle in his winning set, hence the algorithm provides a 2 approximation ($t \neq 0, m$):

$$0 \geq v(t) - ((1 + \alpha)u(m) - u(m - t))$$

$$(1 + \alpha)v(m) - v(t) \geq u(m - t)$$

Which implies that Bob has a non-positive profit for $m - t$ items. Thus from now we assume that $k > 0$. We claim that the optimal solution is $(k, m - k)$. To see that, observe that for every $t \neq 0, m$:

$$v(k) - ((1 + \alpha)u(m) - u(m - k)) \geq v(t) - ((1 + \alpha)u(m) - u(m - t))$$

$$v(k) + u(m - k)) \geq v(t) + u(m - t))$$

We also have to show that the welfare of $(k, m - k)$ is bigger than the welfare of $(m, 0)$:

$$v(k) - ((1 + \alpha)u(m) - u(m - k)) \geq v(m) - u(m)$$

$$v(k) + u(m - k) \geq v(m) + \alpha u(m)$$

From the last inequality we also get that $v(k) \geq \alpha u(m)$ since $u(m - k) \leq v(k)$. Using this fact, we conclude that the approximation ratio that the algorithm provides is at most $\frac{v(k)+u(m-k)}{v(k)} = 1 + \frac{u(m-k)}{v(k)} \leq \frac{u(m)}{\alpha u(m)} = 1 + \frac{1}{\alpha}$. □

## B.3 Fractions Auction

**Definition B.7** *Given constants $0 \leq \alpha_1 \leq ... \leq \alpha_{m-1} \leq 1$, the $(\alpha_{m-1}, \ldots, \alpha_1)$-fractions auction is the following: Bob's payment for $m$ items is $v(m)$, for $t \neq m, 0$ items is $\alpha_t \cdot v(m)$, for $t = 0$ items it is $0$. Alice's payment for $t > 0$ items is $\max\{u(m), \frac{u(m-1)}{\alpha_{m-1}}, \ldots, \frac{u(m-t)}{\alpha_{m-t}}\}$. Alice's payment for $t = 0$ items is $0$.*

**Lemma B.8** *The $(\alpha_{m-1}, \ldots, \alpha_1)$-fractions auction is feasible.*

**Proof:** Suppose that Bob has the bundle of $k$ items in his winning set. We will show that Alice in her winning set a bundle of size at most $m - k$, hence the mechanism is feasible. The bundle of $k$ items has a non-negative profit for Bob:

$$u(k) \geq \alpha_k v(m)$$

Now observe that for every $t > m - k$ the payment of Alice is $\max\{u(m), \frac{u(m-1)}{\alpha_{m-1}}, \ldots, \frac{u(t)}{\alpha_k}\} \geq \frac{u(k)}{\alpha_k}$. For Alice to have $m - k$ items in her winning set (and not the empty bundle) we must therefore have that $v(m) \geq v(m - k) > \frac{u(k)}{\alpha_k}$. However, this is false since $u(k) \geq \alpha_k v(m)$. □

**Lemma B.9** *The $(\alpha_{m-1}, \ldots, \alpha_1)$-fractions auction provides an approximation ratio of $\frac{2}{\alpha_1}$.*

**Proof:** Suppose that Bob is allocated $k > 0$ items. Since the bundle of $k$ items is has a non-negative profit for Bob, we have that $u(k) \geq \alpha_k v(m)$, and this is a lower bound to the welfare of the allocation. The welfare of the optimal allocation is at most $v(m) + u(m)$. Observe that since Bob is allocated $k$ and not $m$ items we have that $u(k) - \alpha_k v(m) \geq u(m) - v(m)$ and therefore $u(m) - u(k) \leq (1 - \alpha_k)v(m)$. The approximation ratio of the algorithm in this case is at least

$$\frac{v(m) + u(m)}{u(k)} \leq \frac{v(m) + u(k) + (1 - \alpha_k)v(m)}{u(k)} \leq \frac{u(k)}{u(k)} + \frac{v(m) + (1 - \alpha_k)v(m)}{u(k)} \leq 1 + \frac{v(m) + (1 - \alpha_k)v(m)}{\alpha_k v(m)}$$

$$\leq \frac{2}{\alpha_k} \leq \frac{2}{\alpha_1}$$

The other case is when Bob is allocated the empty bundle. Let $k$ be the number of items that Alice is assigned. Observe that $k > 0$: since Bob is not allocated any items, we have that for all $t \neq 0, m$, $u(t) \leq \alpha_t \cdot v(m)$ and also $u(m) \leq v(v)$. Thus Alice has a non-negative profit from taking the bundle of $m$ items, therefore $k > 0$. In particular we have that $v(k) \geq \max\{u(m), \frac{u(m-k)}{\alpha_{m-k}}\}$. As in the previous case, we also have that, since Alice prefers $k$ items over $m$ items, $v(m) - \max\{u(m), \frac{u(m-1)}{\alpha_{m-1}}, \ldots, \frac{u(1)}{\alpha_1}\}\alpha_t \leq v(k) - \max\{u(m), \frac{u(m-1)}{\alpha_{m-1}}, \ldots, \frac{u(m-k)}{\alpha_{m-k}}\}$. Thus, $v(m) - v(k) \leq \frac{u(m)}{\alpha_1} - u(m)$.

Again, an upper bound on the value of the optimal allocation is $v(m) + u(m)$. The approximation ratio of the algorithm is at least

$$\frac{u(m) + v(m)}{v(k)} = \frac{v(m)}{v(k)} + \frac{u(m)}{v(k)} \leq \frac{v(k) + (\frac{u(m)}{\alpha_1} - u(m))}{v(k)} + \frac{u(m)}{v(k)} \leq 1 + \frac{u(m)}{\alpha_1 v(k)} \leq 1 + \frac{1}{\alpha_1} \leq \frac{2}{\alpha_1}$$

where the one-before-last inequality is since $u(m) \leq v(k)$, and the last one is since $\alpha_1 \leq 1$. $\square$

## C  Characterization of Scalable 2-Item Auctions

### C.1  Proof of Proposition 4.3

The first direction is trivial. Truthfulness implies the existence of payments $p_1^{(2)}(v(1), v(2))$ and $p_2^{(2)}(v(1), v(2))$ for Bob that depend only on $v(1), v(2)$. So now define $p(r) = p_2^{(2)}(r, 1)$ and $f(r) = p_1^{(2)}(r, 1)/p_2(r, 1)$. Our mechanism on input $(v(1), v(2))$ by definition gives the two-item payment $v(2) \cdot p(v(1)/v(2)) = v(2) \cdot p_2^{(2)}(v(1)/v(2), 1) = p_2^{(2)}(v(1), v(2))$, where the last equality follows from the scalability of $p_2^{(2)}$. Similarly the payment given for one items is $v(2) \cdot p(v(1)/v(2)) \cdot f(v(1)/v(2)) = v(2) \cdot p_2^{(2)}(v(1)/v(2), 1) \cdot p_1^{(2)}(v(1)/v(2), 1)/p_2^{(2)}(v(1)/v(2), 1) = v(2) \cdot p_1^{(2)}(v(1)/v(2), 1) = p_1^{(2)}(v(1), v(2))$, where the last equality follows from the scalability of $p_1^{(2)}$.

### C.2  Separating the high range

In this section we show that for $r \leq \theta_A$ we have that $f(r) \leq \theta_B$. Similarly it follows that for $s \leq \theta_B$ we have that $g(s) \leq \theta_A$. Note that by the previous section $r \leq \theta_A$ if and only if $p(r) > w$, and this last condition is what drives this section.

At this point we separate into two cases, according to whether $r > wf(r)$. We start with the easy case: we show that if the payment for one item is "too high" then we do not get the required approximation ratio.

**Lemma C.1 (case I)** *If $r \leq \theta_A$ and $r \leq wf(r)$ then $f(r) \leq \theta_B$.*

**Proof:** Assume by way of contradiction that $f(r) > \theta_B$ and so $q(f(r)) > w^{-1}$, and for $\epsilon$ small enough consider the input:

| Number of items | Alice's value | Alice's payment | Bob's value | Bob's payment |
|---|---|---|---|---|
| One | $r$ | ? | $f(r)(1-\epsilon)w$ | $f(r)w$ |
| Two | 1 | $w(1-\epsilon)q(f(r)) > 1$ | $(1-\epsilon)w$ | $w$ |

Notice that Bob gets negative utility from taking an item or two items and thus takes nothing. Alice gets negative utility from taking two items so can take at most a single item. The total welfare is thus at most $r$, whereas the social optimum is at least $r + wf(r)$. Since $r \leq wf(r)$ this is a contradiction to better than 2-approximation. □

For the second case we first need to prepare two lemmas and a corollary.

**Lemma C.2 (weak one-side inverse)** *If $r \leq \theta_A$ then for any $\delta > 0$, $wg(f(r) - \delta)q(f(r) - \delta) \geq r$.*

**Proof:** Assume to the contrary $wg(f(r) - \delta)q(f(r) - \delta) < r$ and consider the following input:

| Number of items | Alice's value | Alice's payment | Bob's value | Bob's payment |
|---|---|---|---|---|
| One | $r$ | $wg(f(r)-\delta)q(f(r)-\delta)(1+\epsilon) < r$ | $(f(r)-\delta)w(1+\epsilon)$ | $w(f(r)-\delta)$ |
| Two | 1 | ? | $w(1+\epsilon)$ | $w$ |

Bob takes two items. However, when $\epsilon$ is small enough, Alice gets positive utility from one item so she will take (at least) a single item, contradicting feasibility. □

**Lemma C.3** *For $r > \theta_A$ we have that $r > f(r)p(r)$.*

**Proof:** Consider the input:

| Number of items | Alice's value | Alice's payment | Bob's value | Bob's payment |
|---|---|---|---|---|
| One | $r$ | ? | $f(r)p(r)(1-\epsilon)$ | $f(r)p(r)$ |
| Two | 1 | $q(f(r))p(r)(1-\epsilon)$ | $p(r)(1-\epsilon)$ | $p(r)$ |

Bob has negative utility for either one item or two items. Since $p(r) > w$ and $q(f(r)) \geq w^{-1}$, Alice has negative utility for two items, as long as $\epsilon$ is small enough. Thus the total welfare is at most $r$, whereas the social optimum is at least $r + f(r)p(r)(1-\epsilon)$, so for better than 2-approximation we must have $r > f(r)p(r)$. □

**Corollary C.4** *If $f(r) > \theta_B$ then $f(r) > g(f(r))q(f(r))$.*

**Proof:** This is the previous lemma with the roles of the players switched and with $s = f(r)$. □

We are now ready to handle the second case:

**Lemma C.5 (case II)** *If $r \leq \theta_A$ and $r > wf(r)$ then $f(r) \leq \theta_B$.*

**Proof:** Assume towards contradiction that there exists $\delta > 0$ such that $f(r) - \delta > \theta_B$. Combining the weak one-sided inverse lemma and the previous corollary we have that $f(r) - \delta > g(f(r)-\delta)q(f(r)-\delta) \geq r/w$; Contradiction. □

Which concludes this subsection:

**Lemma C.6** *For $r \leq \theta_A$ we have $f(r) \leq \theta_B$. For $s \leq \theta_B$ we have $q(s) \leq \theta_B$.*

**Proof:** The Case I and Case II lemmas cover all possibilities for $f$; for $g$ the situation is symmetric. □

## C.3 The low and mid range

In this subsection we continue dealing with the range $r < \theta_A$ and show that it splits into two sub-ranges, the low-range $r \leq l_A$, for which $f(r) = \theta_B$, and the mid-range, $l_A < r < \theta_A$, for which $l_B < f(r) < \theta_B$. We get that by proving an interesting-by-its-own property of $f$ in this range: continuity.

**Lemma C.7 (monotone consistency)** *For $r \leq \theta_A$ and $s \leq \theta_B$ we have that $s \leq f(r)$ if and only if $r \leq g(s)$.*

**Proof:** We will prove the only if direction, and the other direction is symmetric. Assume towards contradiction that $r > g(s)$ and for small enough $\epsilon > 0$ consider the input:

| Number of items | Alice's value | Alice's payment | Bob's value | Bob's payment |
|---|---|---|---|---|
| One | $r$ | $g(s)(1+\epsilon) < r$ | $s(1+\epsilon)w$ | $wf(r)$ |
| Two | 1 | ? | $(1+\epsilon)w$ | $w$ |

The utility of Bob from one item is at most $s$ fraction of his utility for two items which is positive, so Bob wins two items. Alice wins at least one item, contradiction to feasibility. □

As a corollary of the lemma, we get a strong version of the weak one-sided inverse lemma:

**Corollary C.8 (one-sided inverse)** *If $r \leq \theta_A$ then $g(f(r)) \geq r$.*

**Proof:** Use the previous lemma with $s = f(r)$. □

**Corollary C.9 (monotonicity)** *$f$ is monotone non-increasing on $r \leq \theta_A$.*

**Proof:** Assume that $f(r) < f(r')$, for some $r' < r \leq \theta_A$. Take $s = f(r')$. First, since $s > f(r)$ we apply the monotone consistency lemma to get $r > g(s)$. Second, since $s \leq f(r')$ we apply the previous monotone consistency lemma again to get $r' \leq g(s)$. Putting these two inequalities together gives $r > r'$, contradiction. □

We show that $f$ and $g$ cannot decrease too quickly, and satisfy a Lipschitz condition.

**Lemma C.10 (Lipschitz condition)** *If $0 \leq r' < r < \theta_A$ then $f(r') - f(r) \leq (r - r') \cdot f(r)/(1-r)$ and if $0 \leq s' < s < \theta_B$ then $g(s') - g(s) \leq (s - s') \cdot g(s)/(1-s)$.*

**Proof:** As usual we will prove for $f$, and the case for $g$ is similar. The required outcome is equivalent to $f(r')/f(r) \leq (1-r')/(1-r)$, so assume towards contradiction that $f(r')/f(r) > \alpha > (1-r')/(1-r)$, and consider the following input:

| Number of items | Alice's value | Alice's payment | Bob's value | Bob's payment |
|---|---|---|---|---|
| One | $r\alpha$ | $g(f(r'))$ | $wf(r')$ | $\alpha w f(r) < wf(r')$ |
| Two | $\alpha$ | 1 | $w$ | ? |

Bob gets positive utility for one item so he takes (at least) an item. Since $\alpha > (1-r')/(1-r) > 1$, taking two items is profitable for Alice and she will take both items as long that is preferable to taking one item, i.e., if $\alpha - 1 > \alpha r - g(f(r'))$. But our assumption that $\alpha > (1-r')/(1-r)$ is equivalent to $\alpha - 1 > \alpha r - r'$, and that implies the previous inequality since, by the one-sided inverse corollary, $g(f(r')) \geq r'$. Thus all together at least three items are allocated. Contradiction. □

**Corollary C.11 (continuity)** *The function $f$ is continuous on $[0, \theta_A)$ and $g$ is continuous on $[0, \theta_B)$.*

**Proof:** The Lipschitz condition implies continuity. □

We are now able to separate the region for which $f(r) = \theta_B$ from that which $f(r) < \theta_B$. Monotonicity implies that the first is a prefix, so let us define:

**Definition C.12** *Let $l_A = \sup\{r \leq \theta_A \mid f(r) = \theta_B\}$ and $l_B = \sup\{s \leq \theta_B \mid g(s) = \theta_A\}$.*

## C.4 The mid range

This is the central part of the proof in which we show that in the middle range $f$ and $g$ are linear, and in fact the payments are identical to those given by weighted VCG.

**Claim C.13 (inverses)** *For $l_A < r < \theta_A$ we have $g(f(r)) = r$.*

**Proof:** The monotone consistency lemma implies that for $s \leq f(r)$ we have $g(s) \geq r$ whereas its contra-positive states that for $s > f(r)$ we have $g(s) < r$. So continuity implies that for $s = f(r)$ we have $g(s) = r$. □

**Corollary C.14 (bijective)** *The function $f(r)$ is bijective from the interval $(l_A, \theta_A)$ to the interval $(l_B, \theta_B)$ and the function $g(s)$ is bijective from the interval $(l_B, \theta_B)$ to the interval $(l_A, \theta_A)$.*

The following lemma shows that the Lipschitz bound above is actually tight and defined a constant derivative for $f$, at least for small enough $r$. The approximation ratio of the mechanism plays a similar role to that of the role the feasibility of the mechanism played in the proof of Lipschitz condition. Hence, from now on to the end of this subsection we fix the guaranteed approximation ratio of the mechanism to be $2/(1+\delta)$, for $0 < \delta \leq 1$.

**Lemma C.15 (differences)** *For every $l_A < r' < r < \theta_A$ and $r \leq (1+\delta)wf(r')(1-r)/(1-r')$ we have $f(r') - f(r) = (r - r') \cdot f(r)/(1-r)$.*

**Proof:** The required outcome is equivalent to $f(r')/f(r) = (1-r')/(1-r)$, so assume towards contradiction that $f(r')/f(r) < \alpha < (1-r')/(1-r)$ (since the other direction was shown to be impossible), and consider the following input:

| Number of items | Alice's value | Alice's payment | Bob's value | Bob's payment |
|---|---|---|---|---|
| One | $r\alpha$ | $g(f(r'))$ | $wf(r')$ | $\alpha w f(r)$ |
| Two | $\alpha$ | $1$ | $w$ | $\alpha w$ |

Bob will not take both items since $\alpha > f(r')/f(r) > 1$, and will not take a single item since $\alpha f(r) > f(r')$. Alice will not take both items if taking one item is preferable, i.e., if $\alpha - 1 < \alpha r - g(f(r')) = \alpha r - r'$ which is equivalent to our assumption that $\alpha < (1-r')/(1-r)$. Thus the total utility obtained is $r\alpha$ whereas the optimum is at least $r\alpha + wf(r')$ in contradiction to $2/(1+\delta)$-approximation since $\alpha < (1-r')/(1-r)$ implies $r\alpha < (1+\delta)wf(r')$. □

The differences lemma gives us an important property of $f$. We now use it to show that in some regions $f$ behaves in a certain linear way.

**Lemma C.16** *For $l_A < r < \theta_A$ with $r < (1+\delta)wf(r)$ we have that $f(r) = c \cdot (1-r)$ for some constant $c > 0$.*

**Proof:** Take $l_a < r' < r$. By the differences lemma $f(r)/(1-r) = f(r')/(1-r')$. Thus we have the required result in an open interval around $r$ (by setting $c = f(r')/(1-r')$, and since this is true for every $r$ and the intervals are overlapping, it must be the same constant $c$ everywhere. $\square$

**Corollary C.17** *For $l_A < r < \theta_A$ with $r(1+\delta) > wf(r)$ we have that $f(r) = 1 - c'r$ for some constant $c' > 0$.*

**Proof:** The symmetric version of the previous lemma for $g$ states that for $s < (1+\delta)w^{-1}g(s)$ we have $g(s) = c''(1-s)$. Use $s = f(r)$ and the fact that $g(f(r)) = r$ to obtain that for $f(r) < (1+\delta)w^{-1}r$ we have $r = g(f(r)) = c''(1-f(r))$, which implies the corollary, for $c' = \frac{1}{c''}$. $\square$

Combining the last lemma and last corollary together we get that $f$ behaves in the mid range like the VCG mechanism:

**Corollary C.18** *For every $l_A < r < \theta_A$ we have that $f(r) = 1 - r$.*

**Proof:** Since $f$ is continuous, the range $r(1+\delta) > wf(r)$ overlaps the range $r < (1+\delta)wf(r)$. Thus for every $r$ in this interval we have that $c(1-r) = f(r) = 1 - c'r$. This implies $c = c' = 1$. $\square$

Which leads us to the conclusion of this subsection:

**Lemma C.19 (summary of subsection)** *For every $1 - \theta_B < r < \theta_A$ we have that $f(r) = 1 - r$. For every $\theta_A < s < \theta_B$ we have that $g(s) = 1 - s$. For every $r < 1 - \theta_B$ we have that $f(r) = \theta_B$. For every $s < 1 - \theta_A$ we have that $g(s) = \theta_A$.*

**Proof:** As $g$ is monotone decreasing, continuous and onto, we must have $\lim_{s \to \theta_B} g(s) = l_A$. The previous corollary allows directly evaluating the limit to be $\theta_B$ and the $l_A = 1 - \theta_B$, which gives the desired result as the statements of the previous corollary as well as the definition of $l_A$ and $l_B$. $\square$

## C.5 The Value of $f(\theta_A)$ and $g(\theta_B)$

B now we have completed the characterization of $f$ and $g$ for the mid and low range. Before handling the high range, we handle the transition points between the mid and high range, i.e., $f(\theta_A)$ and $g(\theta_B)$.

**Lemma C.20** $f(\theta_A) = 1 - \theta_A$ and $g(\theta_B) = 1 - \theta_B$.

We prove the lemma for $f(\theta_A)$ but it symmetrically holds for $g(\theta_B)$. The proof consists of the following two claims:

**Claim C.21** $f(\theta_A) \geq 1 - \theta_A$.

**Proof:** Suppose towards a contradiction that $f(\theta_A) < 1 - \theta_A$. Consider the following instance, where $\delta > 0$:

| Number of items | Alice's value | Alice's payment | Bob's value | Bob's payment |
|---|---|---|---|---|
| One | $\theta_A$ | $(1-\delta)\theta_A$ | $(1-\delta)(\frac{wf(\theta_A)+1-\theta_A}{2})$ | $wf(\theta_A)$ |
| Two | 1 | $(1-\delta)$ | $(1-\delta)w$ | $w$ |

Notice that Alice's payment for one item is indeed $(1-\delta)\theta_A$, since Bob's value for one item is in the low range. Alice is allocated 2 items (her most profitable bundle). We choose $\delta > 0$ such that $(1-\delta)(\frac{wf(\theta_A)+1-\theta_A}{2}) > wf(\theta_A)$, and thus Bob is allocated at least one item. This is a contradiction to the feasibility of the mechanism. □

**Claim C.22** $f(\theta_A) \leq 1 - \theta_A$.

**Proof:** Suppose towards a contradiction that $f(\theta_A) > 1 - \theta_A$. Choose $t$ to be such that $f(\theta_A) > t > 1 - \theta_A$. Consider the following instance:

| Number of items | Alice's value | Alice's payment | Bob's value | Bob's payment |
|---|---|---|---|---|
| One | $\theta_A$ | $(1+\delta)g(t)$ | $(1+\delta)wt$ | $wf(\theta_A)$ |
| Two | 1 | $(1+\delta)$ | $(1+\delta)w$ | $w$ |

If $\delta > 0$, Bob is allocated 2 items. Also notice that $g(t) < \theta_A$, since $t$ is in the mid range. Thus if $\delta$ is small enough taking one item is profitable for Alice. Hence the mechanism is allocating at least 3 items, in contradiction to the feasibility of the mechanism. □

## C.6 The high range

In this section we characterize the high range. We prove that for every $r > \theta_A$, $f(r) = \frac{r}{\theta_A} - r$, and for every $r > \theta_B$, $g(r) = \frac{r}{\theta_B} - r$. The first two claims we prove establish together that $\theta_A \geq wf(\theta_A)$.

**Claim C.23** For every $r > \theta_A$, $r > wf(r)$. Similarly, for every $r > \theta_B$, $\theta_B > g(r)/w$.

**Proof:** We prove only the first part. The proof is similar to the proof of Lemma C.3. Suppose towards a contradiction that $r \leq wf(r)$. Consider the following instance, for $\gamma > 0$:

| Number of items | Alice's value | Alice's payment | Bob's value | Bob's payment |
|---|---|---|---|---|
| One | $r$ | ? | $wf(r) - \gamma$ | $wf(r)$ |
| Two | 1 | $>1$ | $w(1+\gamma)$ | $\frac{wr}{\theta_A}$ |

If $\gamma$ is small enough then Bob is not allocated any items. Alice is allocated at most one item. Thus, when $\gamma, \epsilon$ approach 0 the approximation ratio approaches 2: the solution that allocates one item to each of the bidders has value of $r + \epsilon + wf(r) - \gamma$ whereas the algorithm returns a solution with value at most $r \leq wf(\theta_A)$. □

**Claim C.24** For every $r > \theta_A$, $wf(\theta_A) - wf(r) \leq w(1-\theta_A)$. For every $r > \theta_B$, $w^{-1}g(\theta_B) - w^{-1}g(r) \leq w^{-1}(1-\theta_B)$.

**Proof:** Consider the following instance, where $\epsilon = \frac{r-\theta_A}{1-\theta_A}$:

| Number of items | Alice's value | Alice's payment | Bob's value | Bob's payment |
|---|---|---|---|---|
| One | $r$ | $\theta_A(1-\epsilon)$ | $w(1-\theta_A)(1-\epsilon)$ | $wf(r)$ |
| Two | 1 | $1-\epsilon$ | $w(1-\epsilon)$ | $\frac{wr}{\theta_A}$ |

Alice's profit from both one and two items is positive and equal $\epsilon$. Thus for any slightly bigger value of $\epsilon$ Alice will be allocated two items. However if $wf(r) < w(1 - \theta_A)(1 - \epsilon)$, Bob will be allocated one item. This is a contradiction to the feasibility of the mechanism as at least three items are allocated overall. This implies the claim. □

**Corollary C.25** $\theta_A \geq wf(\theta_A)$. Similarly, $\theta_B \geq f(\theta_B)/w$.

**Proof:** From the last claim we have that as $r > \theta_A$ approaches $\theta_A$, $wf(\theta_A) - wf(r)$ either approaches 0 or is negative. Since we also have that for every $r > \theta_A$, $r > wf(r)$, this implies that $\theta_A \geq wf(\theta_A)$. □

Before giving the precise description of the high range, we need one last claim:

**Claim C.26** Let Bob's valuation be $u = (w + (\theta_A - w(1 - \theta_A)) + \epsilon, \theta_A + \epsilon)$, for some small enough $\epsilon > 0$. Denote Alice's payment for one item by $p$. Then, $p = \theta_A$. Similarly, let Alice's valuation be $v = (\frac{1}{w} + (\theta_B - (1 - \theta_B)/w) + \epsilon, \theta_B + \epsilon)$, for some small enough $\epsilon > 0$. Bob's payment for one item is $\theta_B$.

**Proof:** We prove only the first part; The second part is very similar. Consider the following instance:

| Number of items | Alice's value | Alice's payment | Bob's value | Bob's payment |
|---|---|---|---|---|
| One | $\theta_A$ | $(1+\delta)p$ | $(1+\delta)(\theta_A + \epsilon)$ | $w(1 - \theta_A)$ |
| Two | 1 | $>1$ | $(1+\delta)(w + (\theta_A - w(1 - \theta_A)) + \epsilon)$ | $w$ |

When $\delta = 0$, the profit of Bob from either one item or two items is equal and non-negative (by Claim C.23, since $\theta_A \geq wf(\theta_A)$). Thus, when $\delta > 0$ Bob is allocated two items and Alice is allocated no items. This implies that $p \geq (1+\delta)\theta_A$. When $\delta < 0$ the no bidder is allocated two items. If $\delta < 0$ and $\epsilon > 0$ are small enough, to preserve the approximation ratio each bidder, and in particular Alice, must be allocated one item (since each bidder contributes about half of the value of the solution that allocates one item to each bidder). In this case we therefore have that $p \leq (1+\delta)\theta_A$. Taking $\delta$ to 0 from above and below we get that $p = \theta_A$, as needed. □

We are now about to finish the proof of the characterization by giving a formula for the high range:

**Lemma C.27** Let $r > \theta_A$. Then $f(r) = \frac{r}{\theta_A} - r$. Similarly, for every $r > \theta_B$, $g(r) = \frac{r}{\theta_B} - r$.

**Proof:** We prove only the first statement. Let $1 \geq t \geq \theta_A$. Consider the following instance (Alice's payment for one item is by Claim C.26):

| Number of items | Alice's value | Alice's payment | Bob's value | Bob's payment |
|---|---|---|---|---|
| One | $\theta_A$ | $(1+\delta)\theta_A$ | $(1+\delta)(\theta_A + \epsilon)$ | $wtf(\frac{\theta_A}{t})$ |
| Two | $t$ | $>1$ | $(1+\delta)(w + (\theta_A - w(1 - \theta_A)) + \epsilon)$ | $w$ |

We will show that when $\delta = 0$, for every $1 \geq t \geq \theta_A$, Bob's payments are identical: for one item the payment is always $w(1 - \theta_A)$ and for 2 items it is $w$. This implies, using scalability, that $tf(\frac{\theta_A}{t}) = f(\theta_A) = 1 - \theta_A$. From the last equation we can calculate $f(r)$ for every $1 \geq r \geq \theta_A$: using $t$ such that $r = \frac{\theta_A}{t}$, we have that $\frac{\theta_A}{r}f(r) = 1 - \theta_A$, and therefore $f(r) = \frac{r}{\theta_A} - r$, as needed.

We start by showing that Bob's payment for two items is identical for all $1 \geq t \geq \theta_A$. Using scalability and the formula for the payment for two items, the payment is $w \cdot t \cdot \frac{\theta_A}{t} \cdot \frac{1}{\theta_A} = w$.

We now show that the Bob's payment for one item is the same for all such $t$. Suppose for contradiction that for some $t$, $tf(\frac{\theta_A}{t}) > 1 - \theta_A$. Observe that for small enough $\delta < 0$ the Bob is allocated two items, whereas Alice is allocated one item – a contradiction to the feasibility of the mechanism. Now suppose that for some $t$, $tf(\frac{\theta_A}{t}) < 1 - \theta_A$. In this case fix some small enough $\delta > 0$ and observe that Bob is allocated one item, whereas Alice is allocated no items at all. In this case the value of the solution that allocates one item to each bidder is $\theta_A + (1+\delta)\theta_A$, but the value of the solution the algorithm obtains is only $(1+\delta)\theta_A$. The algorithm provides an approximation ratio that approaches 2 as $\delta$ approaches 0. A contradiction. □

## D  Missing Proofs of Section 5

### D.1  Proof of Proposition 5.3

We prove that the four properties hold. Recall that $(a_1, a_2)$ is the output of $A$ and $(a'_1, a'_2)$ be the output of the induced mechanism.

- **Feasibility:** Since $l_1 + h_2 > m$ if Bob is allocated two items in $A^{l_1,h_1,l_2,h_2}$ then Alice is allocated no items. Similarly, $l_2 + h_1 > m$ so if Alice is allocated two items then Bob is allocated no items. Thus $A^{l_1,h_1,l_2,h_2}$ is feasible if there is a bidder that is allocated two items. Feasibility is obvious when each of the bidders is allocated at most one item in $A^{l_1,h_1,l_2,h_2}$.

- **Truthfulness:** We prove for Alice with valuation $v$. The proof for Bob with valuation $u$ is similar. Observe that $v^{l_1,h_1}(a_1) = v(a'_1)$ and that the payment of Alice is identical in $A$ and in $A^{l_1,h_1,l_2,h_2}$. Hence, the profit of Alice from taking $t$ items in $A$ is identical to her profit from taking $t'$ items in the induced mechanism ($t' = 0$ if $t < l_1$, $t' = 2$ if $t \geq h_1$ and $t' = 1$ otherwise). Thus, since $A$ is a truthful mechanism and Alice is allocated her most profitable bundle in $A$, she is also allocated her most profitable bundle in the induced mechanism. To conclude this proof, observe that since $A$ is truthful the payment of Alice depends only on Bob's valuation.

- **Approximation Ratio:** Observe that for every allocation of 2 items $s' = (s'_1, s'_2)$ there is an allocation for $m$ items $s = (s_1, s_2)$ such that $v(s'_1) + u(s'_2) = v^{l_1,h_1}(s_1) + u^{l_2,h_2}(s_2)$, and vice versa. In particular, the value of the optimal solution in the instance $(v, u)$ and in $(v^{l_1,h_1}, u^{l^1,h^1})$ is the same. Thus, if the allocation of $m$ items $(a_1, a_2)$ provides an approximation ratio of $\alpha$ to the welfare, so does the allocation of 2 items $(a'_1, a'_2)$.

- **Scalability:** $A$ is scalable and thus, for every $\alpha > 0$, $(a_1, a_2) = A(v^{l_1,h_1}, u^{l_2,h_2}) = A(\alpha \cdot v^{l_1,h_1}, \alpha \cdot u^{l_2,h_2})$. Since the output of the induced mechanism, $(a'_1, a'_2)$, depends only on the output of $A$, $(a_1, a_2)$, we have that, for every $\alpha > 0$, $(a'_1, a'_2) = A^{l_1,h_1,l_2,h_2}(v, u) = A^{l_1,h_1,l_2,h_2}(\alpha \cdot v, \alpha \cdot u)$.

### D.2  Proof of Claim 5.5

We prove only the first statement, the second one is proved using symmetric arguments. Consider the payment functions $f_1^{l_1,h_1,l_2,h_2}(v)$ and $f_1^{l_1,h_1,l_2,h'_2}(v)$. These functions must be the same since by Corollary 5.4 they equal $\tilde{f}_{l_1}(v')$, where $v'$ is the $(l_1, h_1)$-extension of $v$. Notice that the equality of these two payment functions implies equality of all the parameters that define the single-item payment: $1 - \theta_B^{l_1,h_1,l_2,h_2} = 1 - \theta_B^{l_1,h_1,l_2,h'_2}$ is the infimum of all $r$ such that the derivative of $f_1^{l_1,h_1,l_2,h_2}((1,r)) = f_1^{l_1,h_1,l_2,h'_2}((1,r))$ is negative. $\theta_A^{l_1,h_1,l_2,h_2} = \theta_A^{l_1,h_1,l_2,h'_2}$ is the infimum of all $r \geq 1$ such that the derivative of $f_1^{l_1,h_1,l_2,h_2}((1,r)) = f_1^{l_1,h_1,l_2,h'_2}((1,r))$ is positive. Finally, to see that $w^{l_1,h_1,l_2,h_2} = w^{l_1,h_1,l_2,h'_2}$, notice that $w^{l_1,h_1,l_2,h_2}\theta_B^{l_1,h_1,l_2,h_2} = f_1^{l_1,h_1,l_2,h_2}((1,0)) = f_1^{l_1,h_1,l_2,h'_2}((1,0)) = w^{l_1,h_1,l_2,h'_2}\theta_B^{l_1,h_1,l_2,h'_2}$. Since we already have that $\theta_B^{l_1,h_1,l_2,h_2} = \theta_B^{l_1,h_1,l_2,h'_2}$ we conclude that $w^{l_1,h_1,l_2,h_2} = w^{l_1,h_1,l_2,h'_2}$.

## D.3 Proof of Claim 5.6

We prove only the first statement, the second one is symmetric. As in the proof of Claim 5.5, we have that the equality $f_2^{l_1,h_1,l_2,h_2}(v)$ and $f_2^{l_1,h_1,l'_2,h_2}(v)$, for every 2-item valuation $v$. This implies equality in all parameters that define the two-items payment. To see that, observe that $f_2^{l_1,h_1,l_2,h_2}((1,0)) = f_2^{l_1,h_1,l'_2,h_2}((1,0))$ and that $f_2^{l_1,h_1,l_2,h_2}((1,1)) = f_2^{l_1,h_1,l'_2,h_2}((1,1))$. The first equality shows that $w^{l_1,h_1,l_2,h_2} = w^{l_1,h_1,l'_2,h_2}$. From the second equality we get that $\frac{w^{l_1,h_1,l_2,h_2}}{\theta_A^{l_1,h_1,l_2,h_2}} = \frac{w^{l_1,h_1,l'_2,h_2}}{\theta_A^{l_1,h_1,l'_2,h_2}}$ and thus $\theta_A^{l_1,h_1,l_2,h_2} = \theta_A^{l_1,h_1,l'_2,h_2}$.

## D.4 Proof of Lemma 5.7

Construct a graph where each node represents an induced mechanism of $A$ with parameters $(l_1, h_1, l_2, h_2)$. Construct the following edges: an edge between a node $(l_1, h_1, l_2, h_2)$ and a node $(l'_1, h'_1, l'_2, h'_2)$ exists if and only if exactly one of the following equalities does not hold: $l_1 = l'_1$, $h_1 = h'_1$, $l_2 = l'_2$, $h_2 = h'_2$.

Notice that by Claims 5.5 and 5.6 if two nodes $(l_1, h_1, l_2, h_2)$ and $(l'_1, h'_1, l'_2, h'_2)$ are connected then $w^{(l_1,h_1,l_2,h_2)} = w^{(l'_1,h'_1,l'_2,h'_2)}$. Thus to prove the lemma it suffices to show that the graph is connected. We will show this by observing that there is a path from every node $(l_1, h_1, l_2, h_2)$ to $(1, m, 1, m)$: starting from $(l_1, h_1, l_2, h_2)$, we first increase $h_A$ and then $h_B$ all the way up to $m$ and then reduce $l_A$ and $l_B$ to 1 while changing one index at a time.

## D.5 Proof of Lemma 5.8

The proof uses the following two claims:

**Claim D.1** *Let $m - 2 \geq l_1 \geq 2$. Then, $\theta_A^{l_1,l_1+1,m-l_1,m-l_1+1} = 1$. Similarly, let $m - 2 \geq l_1 \geq 2$. Then, $\theta_B^{m-l_2,m-l_2+1,l_2,l_2+1} = 1$.*

**Proof:** We prove only the first statement. The second one is symmetric. Consider the two induced mechanisms $A^{l_1,l_1+1,m-l_1,m-l_1+1}$ and $A^{l_1-1,l_1,m-l_1+1,m-l_1+2}$. Let $u'$ be the $(m - l_1, m - l_1 + 1)$-extension of the valuation $(1, 1, 0)$. Notice that $u'$ is also the $(m - l_1 + 1, m - l_1 + 2)$-extension of the valuation $(1, 0, 0)$. Thus we have that $\frac{w}{\theta_B^{l_1-1,l_1,m-l_1+1,m-l_1+2}} = g_{l_1}^{l_1-1,l_1,m-l_1+1,m-l_1+2}(u') = g_{l_1}^{l_1,l_1+1,m-l_1,m-l_1+1}(u') = w\theta_A^{l_1,l_1+1,m-l_1,m-l_1+1}$. Since the $\theta_A$ and $\theta_B$ parameters of all induced mechanisms take values between 0 and 1, we have that $\theta_A^{l_1,l_1+1,m-l_1,m-l_1+1} = \theta_B^{l_1-1,l_1,m-l_1+1,m-l_1+2} = 1$. □

**Claim D.2** *Let $m - 2 \geq l_1 \geq 2$. Then, $\theta_A^{l_1,m,l_2,m} = 1$. Similarly, let $m - 2 \geq l_2 \geq 2$. Then, $\theta_B^{l_1,m,l_2,m} = 1$.*

**Proof:** We prove only the first statement. The second one is symmetric. By Claim D.1 we have that $\theta_A^{l_1,l_1+1,m-l_1,m-l_1+1} = 1$. By applying claim 5.5 twice, first increasing the value of $h_1$ to $m$ and then the value of $h_2$ to $m$ we have that $\theta_A^{l_1,m,m-l_1,m} = 1$. By applying claim 5.6 while changing the third coordinate from $m - l_1$ to $l_2$, we conclude that $\theta_A^{l_1,m,l_2,m} = 1$. □

The lemma follows by applying the last claim twice, once for Alice and once for Bob.

## D.6 Proof of Lemma 5.11

We prove only the first statement. The second one is symmetric. The proof consists of the following two claims.

**Claim D.3** *For each $v$, $f_m(v) \geq wv(m)$.*

**Proof:** Assume towards a contradiction that for some $v$, we have that $f_m(v) = w(v(m) - \epsilon)$, for some $\epsilon > 0$. Consider the instance $(v, u)$ where $u = (wv(m) - w\frac{\epsilon}{2}, 0, \ldots, 0)$. Bob's profit for $m$ items is positive whereas his profit for every $k \neq m$ items is at most 0, since $u(k) = 0$. Thus Bob is allocated $m$ items. On the other hand, since $u$ is $k$-simple, $g_m(u) = v(m) - \frac{\epsilon}{w}$, by Corollary 5.10. Thus the profit of Alice from taking $m$ items is positive, hence Alice is not allocated the empty bundle. In conclusion, more than $m$ items are allocated, a contradiction to the feasibility of the mechanism. □

**Claim D.4** *For each $v$ where $v(1) = 0$, $f_m(v) \leq wv(m)$.*

**Proof:** Assume towards a contradiction that for some $v$, we have that $f_m(v) = w(v(m) + \epsilon)$, for some $\epsilon > 0$. Consider the instance $(v, u)$ where $u = (w(v(m) + \frac{\epsilon}{2}), 0, \ldots, 0)$. Notice that the profit of the Bob is negative for the bundle of all items and therefore his contribution to the welfare is 0. The contribution of Alice to the welfare is 0 too: $u$ is (in particular) a 2-simple valuation. Thus, by Corollary 5.10, $g_t(u) = (v(m) + \frac{\epsilon}{2}) > v(m)$ for every $t \neq m, 1$. In addition, the bundle of all items has a negative profit for Alice. Thus Alice is allocated at most one item, but $v(1) = 0$ (observe that by the monotonicity of the payments, $g_{m-1}(u) \geq g_{m-2}(u) > v(m)$, thus Alice is not allocated $m - 1$ items). In conclusion, the mechanism outputs an allocation with a welfare of 0, a contradiction to the fact that the mechanism provides a bounded approximation ratio. □

## D.7 Proof of Lemma 5.12

We prove only the first statement. The second one is symmetric. Notice that the lemma holds for $k = m$, by Lemma 5.11. Next, towards a contradiction, assume that for some valuation $v$ and some $k$, $f_k(v) > w(v(m) - v(m - k))$ (we will consider the case where $f_k(v) < w(v(m) - v(m - k))$ later). Let $\epsilon = f_k(v) - w(v(m) - v(m - k))$. Let $u$ be the valuation where $u(m) = w(m) + \epsilon/4$, $u(l) = w(v(m) - v(m - k)) + \epsilon/2$, for $k \leq l < m$ and $u(l) = 0$ for every $l < k$. Consider the instance $(v, u)$. Bob is allocated $m$ items since this is its only profitable bundle. Observe that the payment induce by Bob for $m - k$ items is: $w^{-1}(u(m) - u(k)) = w^{-1}(w(m) + \epsilon/4 - w(v(m) - v(m - k)) - \epsilon/2) = v(m - k) - w^{-1}\epsilon/2$. Hence, Alice's profit from taking $m - k$ items is positive and the mechanism allocates more than $m$ items. A contradiction to the feasibility of the mechanism.

Now, towards a contradiction, assume that $f_k(v) < w(v(m) - v(m - k))$. Let $\epsilon = w(v(m) - v(m - k)) - f_k(v)$. Let $u$ be the valuation where $u(m) = w(v(m) - \epsilon/4)$, $u(l) = w(v(m) - v(m - k) - \epsilon/2)$, for $k \leq l < m$ and $u(l) = 0$ for every $l < k$. Consider the instance $(v, u)$. Bob is allocated at least $k$ items. Observe, however, that Alice is allocated more than $m - k$ items since by Corollary 5.10 her payment for every $t$ items, $1 < t \leq m - k$, is $v(m) - \epsilon/4 - (v(m) - v(m - k) - \epsilon/2) = v(m - k) + \epsilon/4$ and since her profit for the bundle of $m$ items is positive. A contradiction to the feasibility of the mechanism.

26